\documentstyle[12pt]{article}
\input epsf.sty
\begin{document}
\begin{center} 
{\Large{\bf
Topological Nature of Anomalous Hall Effect in Ferromagnets
}}
\vskip 1.5cm
{\Large
Masaru {\sc Onoda}$^{\dagger,}$\footnote{E-mail: m.onoda@aist.go.jp} and 
Naoto {\sc Nagaosa}$^{\dagger,\ast,}
$\footnote{E-mail: nagaosa@appi.t.u-tokyo.ac.jp}
}

$^\dagger$Correlated Electron Research Center (CERC),\\
National Institute of Advanced Industrial Science and Technology (AIST),\\
Tsukuba Central 4, Tsukuba 305-8562\\
$^\ast$Department of Applied Physics, University of Tokyo,\\
Bunkyo-ku, Tokyo 113-8656
\end{center}
\begin{center} 
\begin{bf}
Abstract
\end{bf}
\end{center}
The anomalous Hall effect in two-dimensional ferromagnets 
is discussed to be the physical realization of the parity anomaly in 
(2+1)D, and the band crossing points behave as the 
topological singularity in the Brillouin zone.
This appears as the sharp peaks and the sign changes of the 
transverse conductance $\sigma_{xy}$ as a function of the Fermi energy
and/or the magnetization.
The relevance to the experiments including the three dimensional
systems is also discussed.

\newpage
Anomalous Hall effect (AHE) in ferromagnets is a phenomenon 
that there appears the contribution to the 
transverse resistivity $\rho_H$ proportional to the 
magnetization $M$ in addition to the usual Hall effect \cite{smith}:
\begin{equation}
\rho_H = R_0 B + 4 \pi R_s M
\label{eq_rho_H}
\end{equation}
where $B$ is the magnetic field, $R_0$ is the ordinary Hall constant,
and $R_s$ is the anomalous Hall coefficient. 
It has been recognized that this AHE comes from the spin-orbit 
interaction $\lambda$ because some connection 
between the spin and orbital motion is needed 
\cite{kl,smit,l,fukuyama,kondo-noz}. 
Karplus and Luttinger (KL) \cite{kl} was the first
to study theoretically the AHE based on the band theory, 
and recognized that this effect is due to the inter-band 
matrix element of the current operators. An important point 
is that the eigenstates of the Hamiltonian are still the 
Bloch waves $| n {\vec k}\rangle$ ($n$: band index, $\vec k$:
crystal momentum in the 1st Brillouin zone)
even when the spin-orbit interaction is present, namely
the translational symmetry remains.
Therefore $\sigma_{xy}(\omega)$ for noninteracting electrons 
is given by \cite{kubo}
\begin{eqnarray}
\sigma_{xy}(\omega)
&=& i \sum_{n \ne m} \sum_{\vec k}
\frac{f(\varepsilon_m(\vec k) ) - f(\varepsilon_n(\vec k) ) }
{ \varepsilon_n(\vec k) - \varepsilon_m(\vec k)}
\nonumber\\
&&\qquad\qquad\times
\frac{\langle n {\vec k }| J_y | m {\vec k} \rangle 
\langle m {\vec k }| J_x | n {\vec k} \rangle}
{\omega + i \eta + \varepsilon_m(\vec k)-\varepsilon_n(\vec k)}
\label{eq_Kubo}
\end{eqnarray}
where $J_\mu$ is the current operator, $f(x)$ is the 
Fermi distribution function,
and we take the natural unit, i.e. $\hbar=c=1$.
This is the general expression, and KL result can be obtained
by expanding the above Kubo formula to the first order in $\lambda$.
On the other hand, Smit \cite{smit} 
stressed the essential relevance of the impurity scattering,
which disturbs the translational symmetry.
Luttinger \cite{l} reconsidered this problem by taking 
into account the impurity scattering in a systematic way, 
and essentially found that
$
\sigma_{xy} = \sigma_{KL} + \sigma_{L}
$
where $\sigma_{KL}$ is the original result of KL \cite{kl}
while the second term $\sigma_L$ is the new term which is
proportional to the inverse of the impurity scattering potential.
$\sigma_L$  is strongly dependent on the explicit form of 
the impurity scattering, and vanishes for the short range
potential \cite{fukuyama}.
However most of the later theories regards the modification of the 
scattering events due to the spin-orbit interaction 
as the origin of AHE \cite{kondo-noz}.

Very recently, it has been experimentally recognized 
that there are two distinct types of behaviors of AHE
\cite{ong-ye-taguchi-salamon,katsufuji-sato,ru,science,taguchi1,sato1}.
In the conventional one, $\sigma_{xy}$ decreases 
as the temperature $T$ is lowered, 
and the AHE is attributed to the scattering events influenced 
by the spin-orbit interaction (type I).  
In the other case, $\sigma_{xy}$ increases as $T$ is lowered
and remains finite in the limit $T=0$ (type II).
In the type II systems, various anomalous behaviors of AHE are observed. 
For example, $\rho_H$ in thin films of SrRuO$_3$ 
shows the change of magnitude of $10^{-9} \Omega m$
with the sign inversion 
\cite{ru} as a function of the temperature
$T$ below the ferromagnetic transition temperature $T_c$. 
The change of $\sigma_{xy}$ per layer is estimated as $O(e^2/h)$
by using the data of 
the longitudinal resistivity $\rho\sim 10^{-7} \Omega m$ 
and the averaged lattice parameter $a \sim 4\times10^{-10}m$.
(We appropriately recover Planck constant $h$.)
Another example is the pyrochlore ferromagnet 
(Sm$_{1-x}$Ca$_x$)$_2$Mo$_2$O$_7$ 
showing the strong doping dependence of AHE, 
where hole doping of the order of $x=0.1$ causes the factor of 
$\sim $8 change of $\rho_{H}$ in contrast to the almost unchanged 
$\rho_{xx}$ \cite{taguchi1}. Similar phenomenon
is observed also for the temperature dependence of
$\rho_H$ in below 
$T = 1$K ($\ll T_c \cong 100$K) in Nd$_2$Mo$_2$O$_7$ \cite{sato1}.
This huge sensitivity on $E_F$ and/or $T$ can not be understood 
in terms of the conventional theories on the AHE nor the 
change of $R_0$ below $T_c$.
The purpose of this paper is to give a theoretical explanation
for the distinction of type I and type II systems and the 
anomalous behaviors of $\sigma_{xy}$ in type II systems.

On the other hand, the topological nature of $\sigma_{xy}$ in 2D has 
been extensively studied in the context of quantum Hall effect 
under external magnetic field \cite{tknn-kohmoto}. 
The Kubo formula eq.(\ref{eq_Kubo}) for $\omega=0$ can be 
rewritten as
\cite{tknn-kohmoto}
\begin{eqnarray}
\sigma_{xy}(0) 
&=& i \sum_{n, {\vec k} }f(\varepsilon_n(\vec k) )
\nonumber\\
&&\quad\times
\sum_{m\neq n}\frac{\langle n {\vec k }| J_y | m {\vec k} \rangle 
\langle m {\vec k }| J_x | n {\vec k} \rangle 
- (J_x \leftrightarrow J_y)}
{[\varepsilon_n(\vec k)  -  \varepsilon_m(\vec k)]^2}
\nonumber\\
&=& e^2\sum_{n, {\vec k} } 
f(\varepsilon_n(\vec k) ) 
\left[ \nabla_k \times 
{\vec A_n({\vec k})} 
\right]_z
\label{eq_DC}
\end{eqnarray}
where 
$
\vec{A}_{n} ({\vec k}) 
= -i\: \langle n {\vec k} | \nabla_k |n {\vec k}\rangle.
$
When the Fermi energy is in a gap and $T=0$K, 
this integral is transformed to the "vorticity" corresponding 
to the vector potential ${\vec A}_n(\vec k)$, 
and is an integer called Chern number [times $(2\pi)^{-1}$].

 From the viewpoint of the field theory, the appearance of 
$\sigma_{xy}$ in 2D is closely related to the {\it parity anomaly}
in (2+1)D Dirac fermions, where the Chern-Simons term is generated
with the sign depending on that of the fermion mass \cite{jak}.
In the Bloch wave case, this Dirac fermion corresponds to the band 
crossing \cite{nielsen},
and several models have been proposed for the condensed matter realization
of the parity anomaly \cite{anomaly,haldane}. 
Haldane \cite{haldane} recognized that the quantum Hall effect can be
realized as the result of the magnetic ordering and without 
external magnetic field in 2D.
It should be noted here that parity transformation in (2+1)D
is ${\vec r} = (x,y) \to {\vec r}' = (-x,y)$, and this 
symmetry operation changes the sign of the Chern-Simons term.
Therefore the effective Hamiltonian should break 
both ${\rm P}_{(2+1){\rm D}}$- and T-symmetries 
(at least infinitesimally).

Although the physical realization of Haldane's model \cite{haldane}, i.e.,
the complex transfer integrals between next-nearest-neighbor
sites on the honeycomb lattice, is not so easy, 
more realistic model has been proposed in connection to the 
the spin chirality mechanism of AHE \cite{ohgushi}.
The spin chirality 
${\vec S}_i \cdot {\vec  S}_j \times {\vec S}_k$ is basically the 
solid angle subtended by  the  non-coplanar spin configurations, 
and its relevance to the AHE has been  
investigated in manganites at finite temperature \cite{ong-ye-taguchi-salamon}
and in pyrochlore compounds \cite{katsufuji-sato,science}
at zero or low temperatures.
This spin chirality acts as the effective magnetic field
\cite{anderson-wen-lee}, and it has been found that
the geometry of the lattice is crucial for the 
finite $\sigma_{xy}$ in terms of this mechanism \cite{ohgushi}. 
Roughly speaking, the positive and negative fluxes 
in the unit cell should be inequivalent to avoid the cancellation.
The multiband effect due to the existence of 
plural number of atoms in a unit cell is essential, and each band is 
characterized by the  Chern number \cite{tknn-kohmoto}. 
Therefore the two conditions, i.e., non-coplanar spins and
non-trivial geometrical structure, makes this mechanism rather special.
 
In this letter we explore the connection between the AHE
and the parity anomaly in 2D ferromagnets, and found the followings. 
(i) The parity anomaly and the resultant nonzero $\sigma_{xy}$
is not the special case but is the generic phenomenon 
occurring in ${\rm P}_{(2+1){\rm D}}$- and 
T-broken magnetic ordered state.
Our model is the conventional one with the 
spin-orbit interaction $\lambda$ and the
collinear spins on the simple square lattice (perovskite structure).
(ii)  The band crossings occur very often as we change the spin-orbit 
coupling $\lambda$ and/or the magnetization $m_z$. 
Even though the Fermi energy is in general not in a gap,
these band crossings (Dirac fermions) behave as the topological singularities 
and appear as the sharp peaks and/or the sign change of $\sigma_{xy}$ as a 
function of the Fermi energy or the magnetization $m_z$.

\begin{figure}[h]
  \begin{picture}(300,300)
    \put(0,0){\epsfxsize 300pt \epsfbox{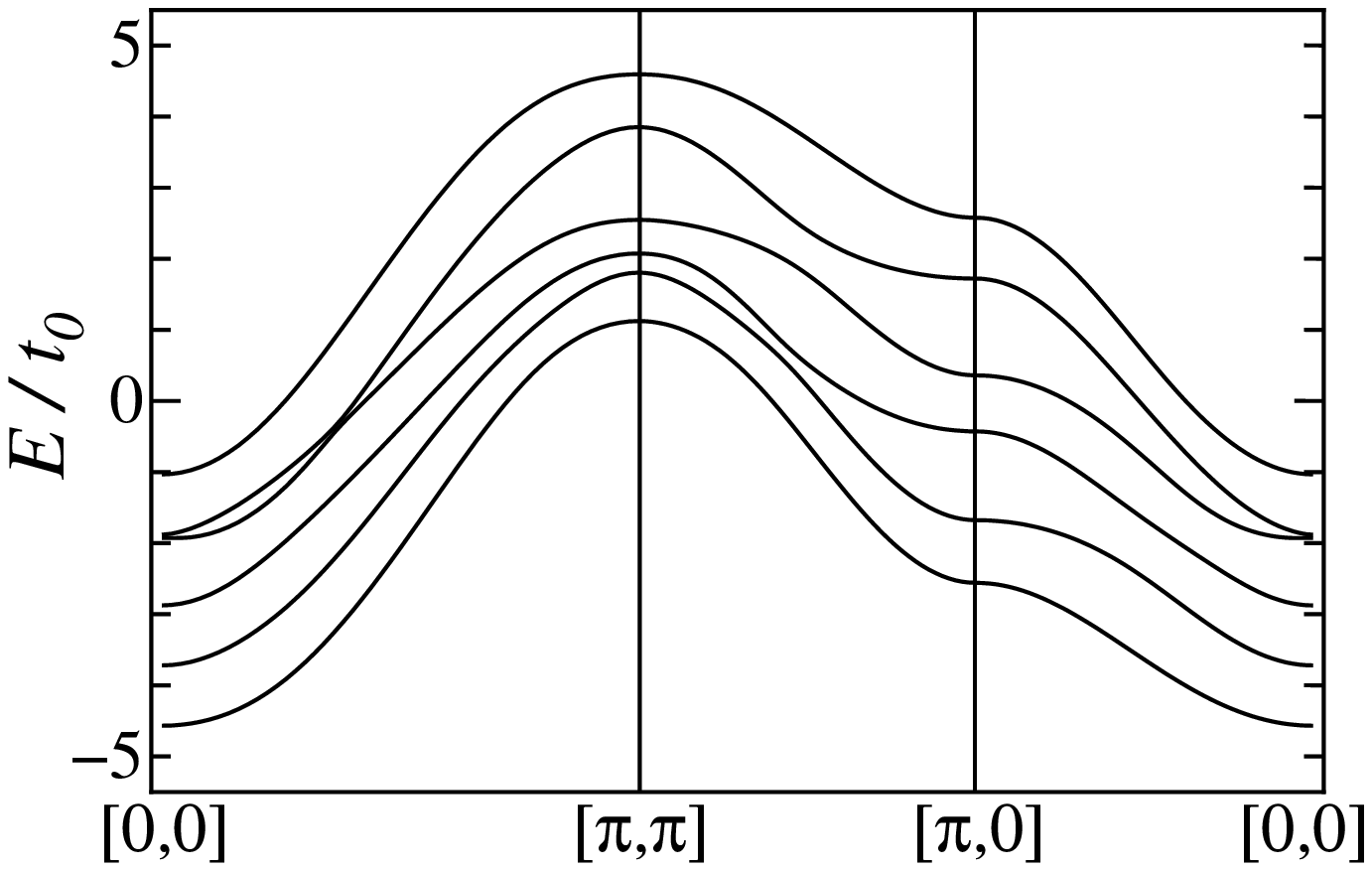}}
  \end{picture}
  \caption{Dispersion of 6 bands for
        $t_1=0.5t_0$, $\lambda=0.4t_0$,
        $Um_z=0.95t_0$.
        The 4th and 5th bands are nearly degenerate
        at $\vec{k}=[0,0]$ and $[\pm\pi/2,\pm\pi/2]$.}
\label{fig_band}
\end{figure}

We consider a model with triply degenerate $t_{2g}$ orbitals with spin-orbit 
interaction on  the 2D square lattice.
\begin{eqnarray}
&&H = -\sum_{i, \sigma} t_0 [
c^\dagger_{xy\: \sigma}(i) c_{xy\:\sigma}(i+x)
+ c^\dagger_{xy\: \sigma}(i) c_{xy\:\sigma}(i+y)
\nonumber\\
&&\quad\quad
+ c^\dagger_{yz\:\sigma}(i) c_{yz\:\sigma}(i+y)
+ c^\dagger_{zx\:\sigma}(i) c_{zx\:\sigma}(i+x)
+ \rm{h.c.} ]
\nonumber \\
&&+ \sum_{i, \sigma, \pm} \pm t_1 
[ c^\dagger_{xy \:\sigma}(i) c_{zx\:\sigma} (i \pm y)
+ c^\dagger_{xy \:\sigma}(i) c_{yz\:\sigma} (i \pm x)
+ \rm{h.c.}]
\nonumber \\
&&+ H_{SO} + H_{\rm el-el}
\end{eqnarray}
where the transfer integrals are determined by the oxygen $p$-orbitals
between the two transition metal ions.
$t_0$ is  nonzero even for 
the perfect perovskite structure, while $t_1$
becomes nonzero due to the shift of the oxygen atoms
out of plane in the positive (negative) $z$-direction.
This will cause the Dzyaloshinsky-Moriya interaction \cite{dm}
when the spin-orbit interaction $H_{SO}$ and 
the transfer integrals are combined, because the 
space inversion symmetry at the middle of the bond is broken.
Our definition of ${\rm P}_{(2+1){\rm D}}$ here is for the 
effective 2D Hamiltonian, and does not include the transformation 
on the orbital and spin degrees of freedom.
The $t_1$-term breaks 
this  ${\rm P}_{(2+1){\rm D}}$-symmetry.
If the shift of the oxygen atom is alternating, 
the sign $\pm$ in the $t_1$ term is missing
and ${\rm P}_{(2+1){\rm D}}$-symmetry is unbroken.
In this case, the topological effect described below is missing, 
because the gauge flux density 
$b_n(\vec{k})=[ \nabla_k \times {\vec A_n({\vec k})}]_z$ 
in eq.(\ref{eq_DC}) has the symmetry as
$b_n(-k_x, k_y)=-b_n(k_x, k_y)$.
(Note that $b_n(\vec{k})$ itself is not zero.)
We approximate $H_{SO}$
by taking the matrix elements of the angular momentum
$\vec{\ell}$ in the space of $t_{2g}$ orbitals as
\begin{equation}
H_{SO} = \lambda \sum_{i} \sum_{\alpha,\beta}
\sum_{\sigma, \sigma'}  
c^\dagger_{\alpha \sigma}(i) 
\vec{\ell}_{\alpha\beta}
\cdot\vec{\tau}_{\sigma\sigma'}
c_{\beta \sigma'}(i)
\end{equation}
where $\vec{\tau}$ are Pauli matrices and
$\alpha,\beta = xy,yz,zx$.
Although $H_{SO}$ is the on-site interaction, it can 
produce the effective gauge flux combined with the transfer 
integrals. The  electron-electron interaction
$H_{\rm el-el}$ induces the ferromagnetic ordering and 
its details are not relevant to the present study.
We take the mean field approximation as
\begin{equation}
H_{\rm el-el} = - \frac{U}{2}m_z \sum_{i, \alpha, \sigma} \sigma
c^\dagger_{\alpha \sigma}(i) c_{\alpha \sigma}(i),
\end{equation}
where the ferromagnetic moment is along the $z$-direction.
Now the $ 6 \times 6$ Hamiltonian matrix for each $\vec k$-point 
can be easily  diagonalized, and 6 bands are obtains. 
Figure \ref{fig_band} shows the energy dispersion obtained
for a set of parameters given in the caption. 
The near-degeneracies of the two bands are
found at ${\vec k} = [0,0]$ and $[\pm\pi/2,\pm\pi/2]$.

\begin{figure}[h]
  \begin{picture}(300,400)
    \put(0,0){\epsfxsize 300pt \epsfbox{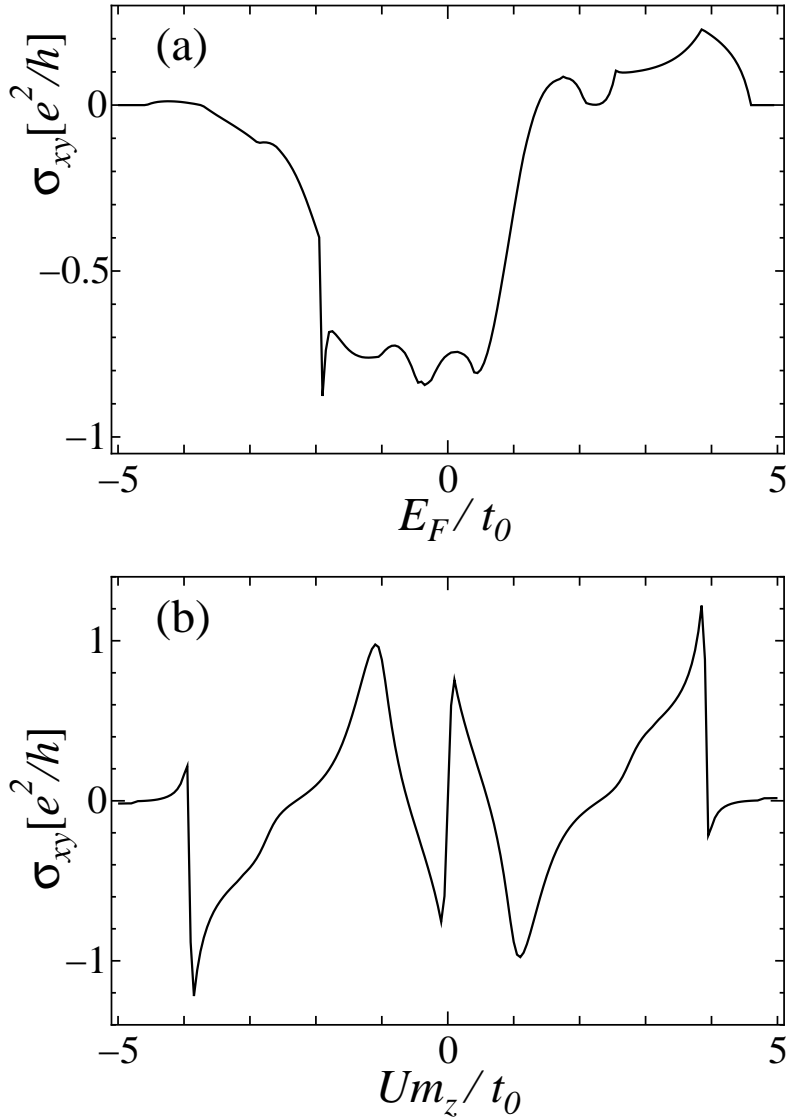}}
  \end{picture}
  \caption{(a) Transverse conductivity $\sigma_{xy}$ 
        as a function of the Fermi energy $E_F$ with
        the parameter set
        $t_1=0.5t_0$, $\lambda=0.4t_0$, $Um_z=0.95t_0$.
        The sharp peak near $-2t_0$ is due to
        the near-degeneracy point at $\vec{k}=[0,0]$
        between the 4th and 5th bands 
        (see Fig.\ref{fig_band}), and, in other words, 
        due to the sharp peak in the gauge flux density 
        in $\vec{k}$-space (see Fig.\ref{fig_flux}).
        (b) Transverse conductivity $\sigma_{xy}$ 
        as a function of the magnetization $m_z$ [times $U$]
        with $t_1=0.5t_0$, $\lambda=0.4t_0$, $E_F=0$.}
\label{fig_sigma_xy}
\end{figure}

Fig.\ref{fig_sigma_xy}(a) shows the typical example of $\sigma_{xy}$
as a function of $E_F$ for the same set of parameters 
as in Fig.\ref{fig_band}, while Fig.\ref{fig_sigma_xy}(b) shows 
the variation as a function of the magnetization
$m_z$ with the fixed $E_F = 0$.
Both of them vary rather rapidly with the magnitude of $e^2/h$
and have sharp peaks.
To analyze this behavior, we first define the Chern number 
$Ch_n$ for each band $n$ as
$Ch_n  = \sum_{ {\vec k} \: \in
{\rm \:1st  \:BZ} }  
2\pi [\nabla_k \times \vec{A}_n({\vec k})]_z.
$
For the parameter set of Fig.\ref{fig_band}, 
the $Ch_n$'s are
(-1, -2, 3, -4, 5, -1)
from the lower to higher energy band.
Namely the spin-orbit interaction gives the 
non-trivial topological structure to the Bloch 
bands in the ferromagnets as in the case of quantum Hall system.

\begin{figure}[h]
  \begin{picture}(320,320)
    \put(0,0){\epsfxsize 320pt \epsfbox{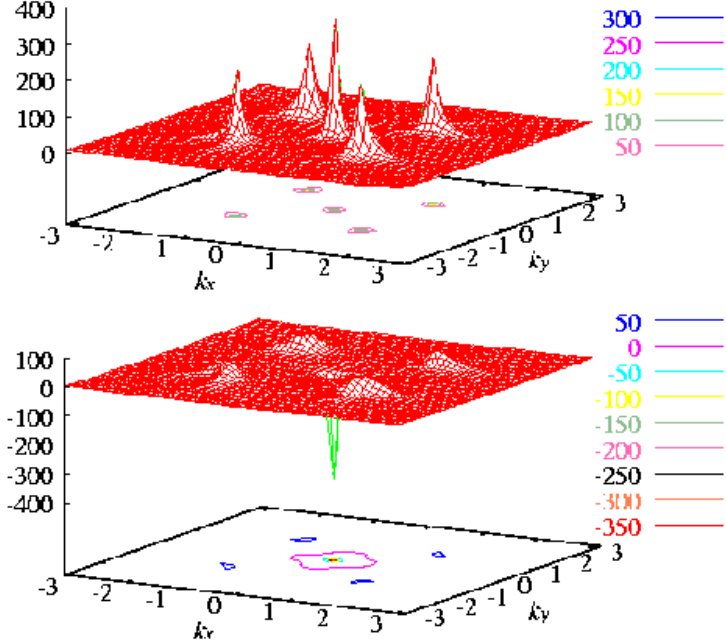}}
  \end{picture}
  \caption{Gauge flux density in $\vec{k}$-space
        of the 5th band with
        the parameter sets
        $t_1=0.5t_0$, $\lambda=0.4t_0$,
        $Um_z=0.95t_0$ for the upper, and
        $Um_z=1.05t_0$ for the lower.
        The transfer of $Ch_n$ occurs
        between the 4th and 5th bands
        at $(U m_z)_c\sim 1.0t_0$. 
        5 peaks are found in both parameter sets.
        However, the transfer occurs only 
        at $\vec{k}=[0,0]$ in this case.}
\label{fig_flux}
\end{figure}

The next question is how the topological 
number appears in $\sigma_{xy}$. In the absence of the gap, 
one has to consider the distribution of 
$[ \nabla_k \times \vec{A}_n({\vec k}) ]_z $
in the $\vec k$-space.
In Fig.\ref{fig_flux} shown 
$[ \nabla_k \times \vec{A}_n({\vec k}) ]_z $
again for the parameter set in Figs.\ref{fig_band} 
and \ref{fig_sigma_xy}(a), where sharp peaks near ${\vec k} = [0, 0]$ 
and ${\vec k} = [\pm\pi/2, \pm\pi/2]$ are clearly seen. 
This behavior can be understood as follows.
When the band degeneracy occurs at $\vec{k} = \vec{k}_0$
for $u = U m_z = u_c$ or $\lambda = \lambda_c$, 
the electrons near this point can be described 
by the generalized Dirac fermions as 
\begin{eqnarray}
&&H \cong \int \frac{d^2 k}{(2\pi)^2}\: 
\psi^{\dagger}(\vec{k}){\cal H}(\vec{k})\psi(\vec{k}),
\quad 
\vec{\kappa} = \frac1{k_D}(\vec{k}-\vec{k}_0),
\nonumber\\
&&{\cal H}(\vec{k}) =
\left[\matrix{
V(\vec{k})+mv^2 & vk_D(\kappa_x-i\kappa_y)^p \cr
vk_D(\kappa_x+i\kappa_y)^p & V(\vec{k})-mv^2 \cr
}\right],
\label{eq_Dirac}
\end{eqnarray}
where $\psi(\vec{k})$ is the two component spinor corresponding 
to the two crossing or touching bands,
$v$ and $k_D$ are positive constants,
and $V(\vec{k})$ is a function of $\vec{k}$.
The usual Dirac fermion corresponds to the case $p=1$.
By diagonalizing ${\cal H}(\vec{k})$, we get the dispersion 
around $\vec{k}_0$ as
$
\varepsilon_{\pm}(\vec{k})
=\pm\sqrt{(vk_D\kappa^p)^2+(mv^2)^2}+V(\vec{k}),
$
where the sign $+/-$ represents the upper/lower band.
The current operator is given by 
$
\vec{J}=\int d^2k/(2\pi)^2\psi^{\dagger}(\vec{k})
\nabla_{k}{\cal H}(\vec{k})\psi(\vec{k})
$.
The mass $m$ is a continuous function of $(\lambda, u)$,
and can changes sign across the critical lines 
in $(\lambda, u)$-plane.
There the transfer of $Ch_n$ occurs between 
these two bands\cite{trans}.
Near this criticality, i.e., when the mass $|m|$ is small,
the gauge flux density 
$b_{\pm}({\vec k})
=[ \nabla_k \times \vec{A}_{\pm}({\vec k}) ]_z$ is peaked 
near the degeneracy point as
\begin{eqnarray}
b_{\pm}({\vec k}) 
&\cong& i\:\frac{\langle\pm,\vec{k}|J_y|\mp,\vec{k}\rangle
\langle\mp,\vec{k}|J_x|\pm,\vec{k}\rangle
-(J_x \leftrightarrow J_y)}
{[\varepsilon_{\pm}(\vec{k})-\varepsilon_{\mp}(\vec{k})]^2}
\nonumber\\
&=& \pm \frac{(pv\kappa^{p-1})^2 mv^2}
{2[(vk_D\kappa^p)^2+ (mv^2)^2]^\frac{3}{2}}.
\label{flux}
\end{eqnarray}
The flux density $b_{\pm}({\vec k})$ 
contributes $\pm \frac{p}{2}\:{\rm sgn}(m)$ 
to the $Ch_n$ of each band.

The transfer of $Ch_n$ has been studied
in connection with the usual Dirac fermion, 
i.e., $p=1$, mainly in the context of 
the integer quantum Hall effect\cite{hatugai}.
However, in the present model, we have actually found also 
its generalizations  given by eq.(\ref{eq_Dirac}).
For example, the type of $p=2$ is seen with the parameter set
($t_1=0.5t_0$, $\lambda\sim 0$, $Um_z=5t_0$),
and the transfer of $Ch_n$ at each degeneracy point 
is $\pm 2$ when $\lambda$ is changed across zero.
Each band touches the next band with
the energy dispersion $E(\vec{k})\propto \vec{k}^2$. 
Another remark here is that the band crossing and 
the transfer of $Ch_n$ occurs frequently. 
It occurs at least 6 times for the parameter set 
($t_1=0.5t_0$, $\lambda=0.4t_0$)
when $Um_z$ is changed from $0$ to $5t_0$.
It occurs more often  if we additionally change $\lambda$.

Some remarks are now in order.
One is on the effect of disorder, which has been neglected thus far.
It is well known that the extended states in 2D is unstable against
the infinitesimal disorder in the absence of magnetic field.
On the other hand, the topological $Ch_n$ should be quantized and 
stable as far as the impurity scattering strength is weak enough 
compared with the gap of the Dirac fermions. This is the basis of 
the existence of the extended states and plateau of $\sigma_{xy}$ 
in the integer QHE, and should work here also \cite{qhe}. 
Therefore we expect that the random potential
realizes the quantum Hall effect in 2D ferromagnets.

Second remark is on the three dimensional case.
The crossing of bands also plays an important role here, but 
the singularity is more mild. 
According to the analysis of Dirac fermions in 3D, 
the peak of $\sigma_{xy}$ near the critical
point is of the order of $|m|$ and the width is again $|m|$.
The magnitude of $\sigma_{xy}$ is of the order of $|m|/t_0$ 
in this case, but it varies over the energy scale of $t_0$
except near the singularity.
Near the singularity, $\sigma_{xy}$ changes rapidly with the
energy scale of $|m|$ while the magnitude remains of the 
order of $|m|/t_0$.

Now the experimental relevance of the present results 
is briefly discussed. 
The most significant consequence of the topological nature
of AHE is its sensitivity to the Fermi energy $E_F$ and/or
the magnetization, in contrast to the diagonal conductivity 
$\sigma_{xx}$ \cite{tokura}.
This naturally explains the anomalous features of AHE in 
thin films of SrRuO$_3$ \cite{ru} and the single
crystal of pyrochlore Mo oxides \cite{taguchi1,sato1} 
mentioned on the introduction.
Especially it has been known that the topological mechanism for AHE
is relevant in the latter material \cite{science}.
The 2D system is even more interesting, where the 
variation in $\sigma_{xy}$ of the order of $e^2/h$,
and with the impurity scattering the quantum Hall effect 
can be also expected. 
It is noted here that the conventional formula eq.(\ref{eq_rho_H}) 
does not work in this case, and $\sigma_{xy}$ should be plotted 
instead. In conclusions, we have studied the anomalous Hall effect
from the viewpoint of the topology and parity/chiral anomaly. 
When both the parity and time-reversal symmetries are broken, 
the AHE survives even at $T=0$ where the periodicity is recovered
( type II). In this case, the band crossing behaves as the 
topological singularity causing the anomalous behavior of 
$\sigma_{xy}$ as a function of the temperature $T$, the 
magnetization $m_z$, and/or the Fermi energy $E_F$. 

The authors acknowledges Y.~Tokura, H.~Fukuyama, and Y.~Taguchi,  
for fruitful discussions. N.~N.  is supported by Priority Areas 
Grants and Grant-in-Aid for COE research from the Ministry of 
Education, Science, Culture and Sports of Japan,


\end{document}